\documentclass[twocolumn]{aastex62}
\usepackage{graphicx}
\usepackage{amssymb}
\usepackage[caption=false]{subfig}
\usepackage{booktabs}
\usepackage{xcolor}

\begin{document}

\title{Lyman-$\alpha$ in the GJ 1132 System: Stellar Emission and Planetary Atmospheric Evolution}

\author[0000-0002-8961-0352]{William C. Waalkes}
\altaffiliation{NSF Graduate Research Fellow}
\affiliation{Department of Astrophysical \& Planetary Sciences, 391 UCB 2000 Colorado Ave, Boulder, CO 80309, USA}

\author[0000-0002-3321-4924]{Zachory Berta-Thompson}
\affiliation{Department of Astrophysical \& Planetary Sciences, 391 UCB 2000 Colorado Ave, Boulder, CO 80309, USA}

\author[0000-0002-9148-034X]{Vincent Bourrier}
\affiliation{Observatoire Astronomique de l'Universit\'e de Gen\`eve, 51 chemin des Maillettes, 1290 Versoix, Switzerland}

\author[0000-0003-4150-841X]{Elisabeth Newton}
\affiliation{Department of Physics and Astronomy, Dartmouth College, Hanover NH 03755}

\author[0000-0001-9704-5405]{David Ehrenreich}
\affiliation{Observatoire Astronomique de l'Universit\'e de Gen\`eve, 51 chemin des Maillettes, 1290 Versoix, Switzerland}

\author[0000-0002-1337-9051]{Eliza M.-R.\ Kempton}
\affiliation{Department of Astronomy, University of Maryland, College Park, MD 20742, USA}
\affiliation{Department of Physics, Grinnell College, 1116 8th Avenue, Grinnell, IA 50112, USA}

\author[0000-0002-9003-484X]{David Charbonneau}
\affiliation{Center for Astrophysics | Harvard-Smithsonian, 60 Garden St., Cambridge, MA 02138, USA}

\author{Jonathan Irwin}
\affiliation{Center for Astrophysics | Harvard-Smithsonian, 60 Garden St., Cambridge, MA 02138, USA}

\author[0000-0001-7730-2240]{Jason Dittmann}
\altaffiliation{51 Pegasi b Postdoctoral Fellow}
\affiliation{MIT, 77 Massachusetts Avenue, Cambridge, MA 02139, USA}

\begin{abstract}
GJ 1132b, which orbits an M dwarf, is one of the few known Earth-sized planets, and at 12 pc away it is one of the closest known transiting planets. Receiving roughly 19x Earth's insolation, this planet is too hot to be habitable but can inform us about the volatile content of rocky planet atmospheres around cool stars. Using Hubble STIS spectra, we search for a transit in the Lyman-$\alpha$ line of neutral hydrogen (Ly$\alpha$). If we were to observe a deep Ly$\alpha$ absorption signature, that would indicate the presence of a neutral hydrogen envelope flowing from GJ 1132b. On the other hand, ruling out deep absorption from neutral hydrogen may indicate that this planet does not have a detectable amount of hydrogen loss, is not losing hydrogen, or lost hydrogen and other volatiles early in the star's life. We do not detect a transit and determine a 2-$\sigma$ upper limit on the effective envelope radius of 0.36 R$_*$ in the red wing of the Ly$\alpha$ line, which is the only portion of the spectrum we detect after absorption by the ISM. We analyze the Ly$\alpha$ spectrum and stellar variability of GJ1132, which is a slowly-rotating 0.18 solar mass M dwarf with previously uncharacterized UV activity. Our data show stellar variabilities of 5-22\%, which is consistent with the M dwarf UV variabilities of up to 41\% found by \citet{Loyd2014}. Understanding the role that UV variability plays in planetary atmospheres is crucial to assess atmospheric evolution and the habitability of cooler rocky exoplanets.
\end{abstract}


\section{Introduction}

The recent discoveries of terrestrial planets orbiting nearby M dwarfs \citep{Gillon2017, Berta-Thompson2015, Dittmann2017, Bonfils2018, Ment2018} provide us with the first opportunity to study small terrestrial planets outside our solar system, and observatories such as the Hubble Space Telescope allow us to analyze the atmospheres of these rocky exoplanets. Additionally, it is important that we learn as much as we can about these planets as we prepare for atmospheric characterization with the James Webb Space Telescope \citep{Deming2009,Morley2017}. JWST will provide unique characterization advantages due to its collecting area, spectral range, and array of instruments that allow for both transmission and emission spectroscopy \citep{Beichman2014}.  

M dwarfs have been preferred targets for studying Earth-like planets due to their size and temperature which allow for easier detection and characterization of terrestrial exoplanets. However, the variability and high UV-to-bolometric flux ratio of these stars makes habitability a point of contention \citep[e.g.,][]{Shields2016, Tilley2017}. It is currently unknown whether rocky planets around M dwarfs can retain atmospheres and liquid surface water or if UV irradiation and frequent flaring render these planets uninhabitable \citep[e.g.,][]{Scalo2007,Hawley2014,Luger2015,Bourrier2017}. On the contrary, UV irradiation may boost the photochemical synthesis of the building blocks of life \citep[e.g.,][]{2018SciA....4R3302R}. We must study the UV irradiation environments of these planets, especially given that individual M stars with the same spectral type can exhibit very different UV properties \citep[e.g.,][]{Youngblood2017}, and a lifetime of UV flux from the host star can have profound impacts on the composition and evolution of their planetary atmospheres.

\begin{table*}[t!]
\begin{center}
\begin{tabular}{ccc}
\toprule
   
    Parameter & Value & Source \\
    \midrule
    \textbf{GJ 1132} \\
    Mass [M$_\Sun$] & 0.181 $\pm$~0.019 & \citet{Berta-Thompson2015}\\
    Radius [R$_\Sun$] &  0.2105$^{+0.0102}_{-0.0085}$
& \citet{Dittmann2017}\\
    Distance [pc] & 12.04 $\pm$~0.24& \citet{Berta-Thompson2015}\\
    Radial Velocity [km~s$^{-1}$] &  35.1 $\pm$~0.8 & \citet{Bonfils2018}\\
    \textbf{GJ 1132b} \\
    Mass [M$_\Earth$] & 1.66 $\pm$~0.23 & \citet{Bonfils2018}\\
    Radius [R$_\Earth$] & 1.13 $\pm$~0.02& \citet{Dittmann2017}\\
    Semi-major Axis, \textit{a} [AU] &  0.0153 $\pm$~0.0005 & \citet{Bonfils2018}\\
    Period [days] & 1.628931 $\pm$~0.000027 & \citet{Bonfils2018}\\
    Epoch [BJD TDB] & 2457184.55786 $\pm$~0.00032 & \citet{Berta-Thompson2015}\\
    $\frac{a}{R_*}$ & 16.54$^{+0.63}_{-0.71}$
 & \citet{Dittmann2017}\\
    \textit{i} (degrees) &  88.68$^{+0.40}_{-0.33}$
& \citet{Dittmann2017}\\
    Surface Gravity [m~s$^{-2}$] & 12.9 $\pm$~2.2& \citet{Bonfils2018}\\
    Equilibrium Temperature, T$_{\rm{eq}}$ [K]: & & \\
    Bond Albedo = 0.3 (Earth-like) & 529 $\pm$~9 & \citet{Bonfils2018} \\
    Bond Albedo = 0.75 (Venus-like) & 409 $\pm$~7 & \citet{Bonfils2018} \\
    
\bottomrule
\end{tabular}
\caption{GJ 1132 system parameters.\label{tab:system}}
\end{center}
\end{table*}

One aspect of terrestrial planet habitability is volatile retention, including that of water in the planet's atmosphere. One possible pathway of evolution for water on M dwarf terrestrial worlds is the evaporation of surface water and subsequent photolytic destruction of H$_2$O into H and O species \citep[e.g.,][]{Bourrier2017,Jura2004}. The atmosphere then loses the neutral hydrogen while the oxygen is combined into O$_2$/O$_3$ and/or resorbed into surface sinks \citep[e.g.,][]{Wordsworth2013, Tian2015, Luger2015, Shields2016, Ingersoll1969}. In this way, large amounts of neutral H can be generated and subsequently lost from planetary atmospheres. Studies have shown O$_2$ and O$_3$ alone to be unreliable biosignatures for M dwarf planets because they possess abiotic formation mechanisms \citep{Tian2013}, though they are still important indicators when used with other biomarkers \citep[see][]{Meadows2018}. Understanding atmospheric photochemistry for terrestrial worlds orbiting M dwarfs is critical to our search for life.

\subsection{Prior Work}

\citet{Kulow2014} and \citet{Ehrenreich2015} discovered that Gliese 436b, a warm Neptune orbiting an M dwarf, has a 56.3$\pm$3.5\% transit depth in the blue-shifted wing of the stellar Ly$\alpha$ line. \citet{Lavie2017} further studied this system to solidify the previous results and verify the predictions made for the structure of the outflowing gas made by \citet{Bourrier2016}. For planets of this size and insolation, atmospheric escape can happen as a result of the warming of the upper layers of the atmosphere, which expand and will evaporate if particles begin reaching escape velocity \citep[e.g.,][]{Vidal-Madjar2003,Lammer2003,Murray-Clay2009}.

\begin{figure}[b!]
\centering
\vspace{0.5cm}
\includegraphics[width=0.48\textwidth]{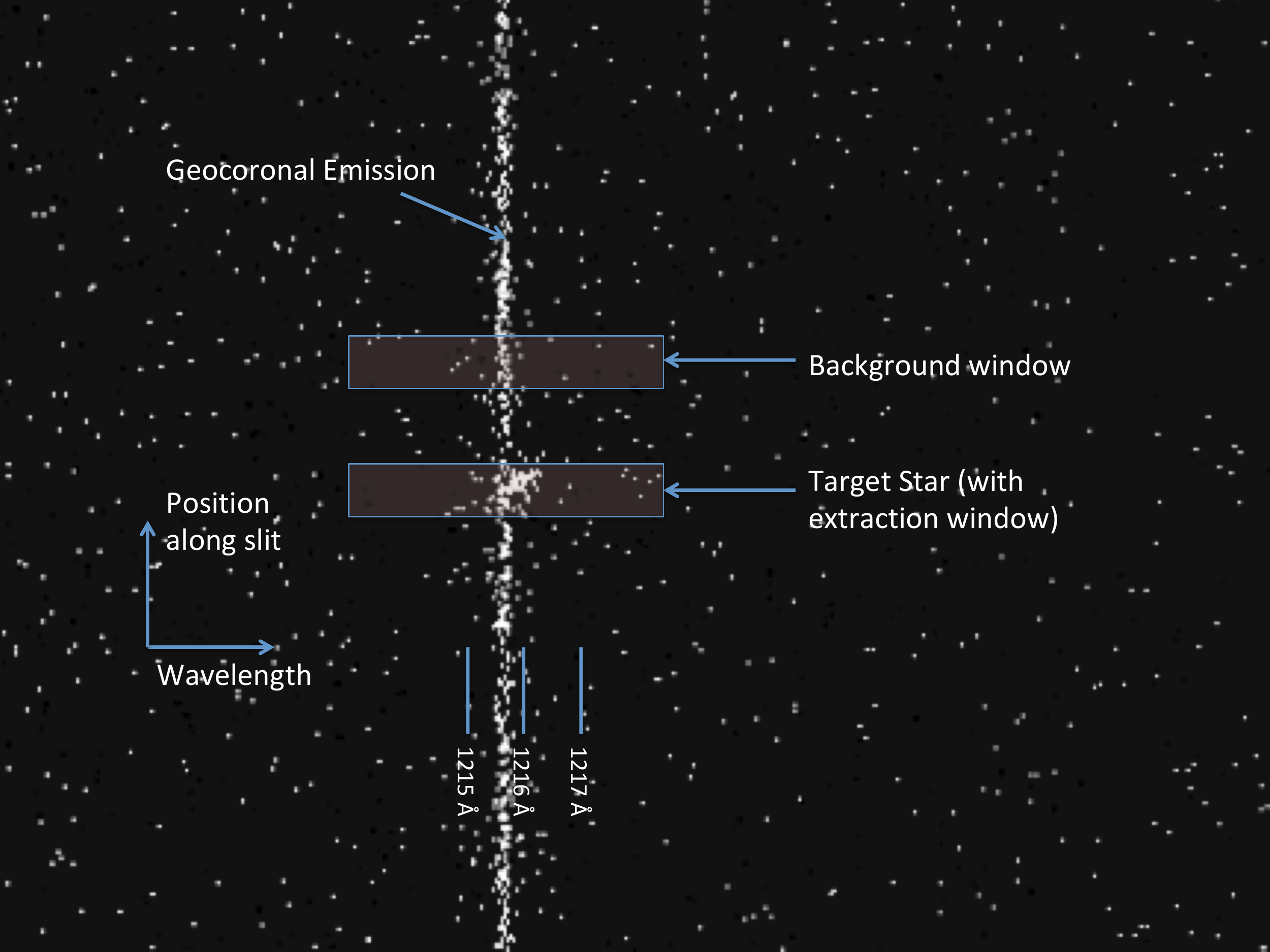}
\caption{Image of a STIS x2d spectrum. Geocoronal Ly$\alpha$ is shown as a long vertical line while the GJ 1132 Ly$\alpha$ emission is shown in the center.\label{x2d}}
\end{figure}

\begin{figure*}[t!]
\centering
\subfloat[]{\includegraphics[width=0.49\textwidth]{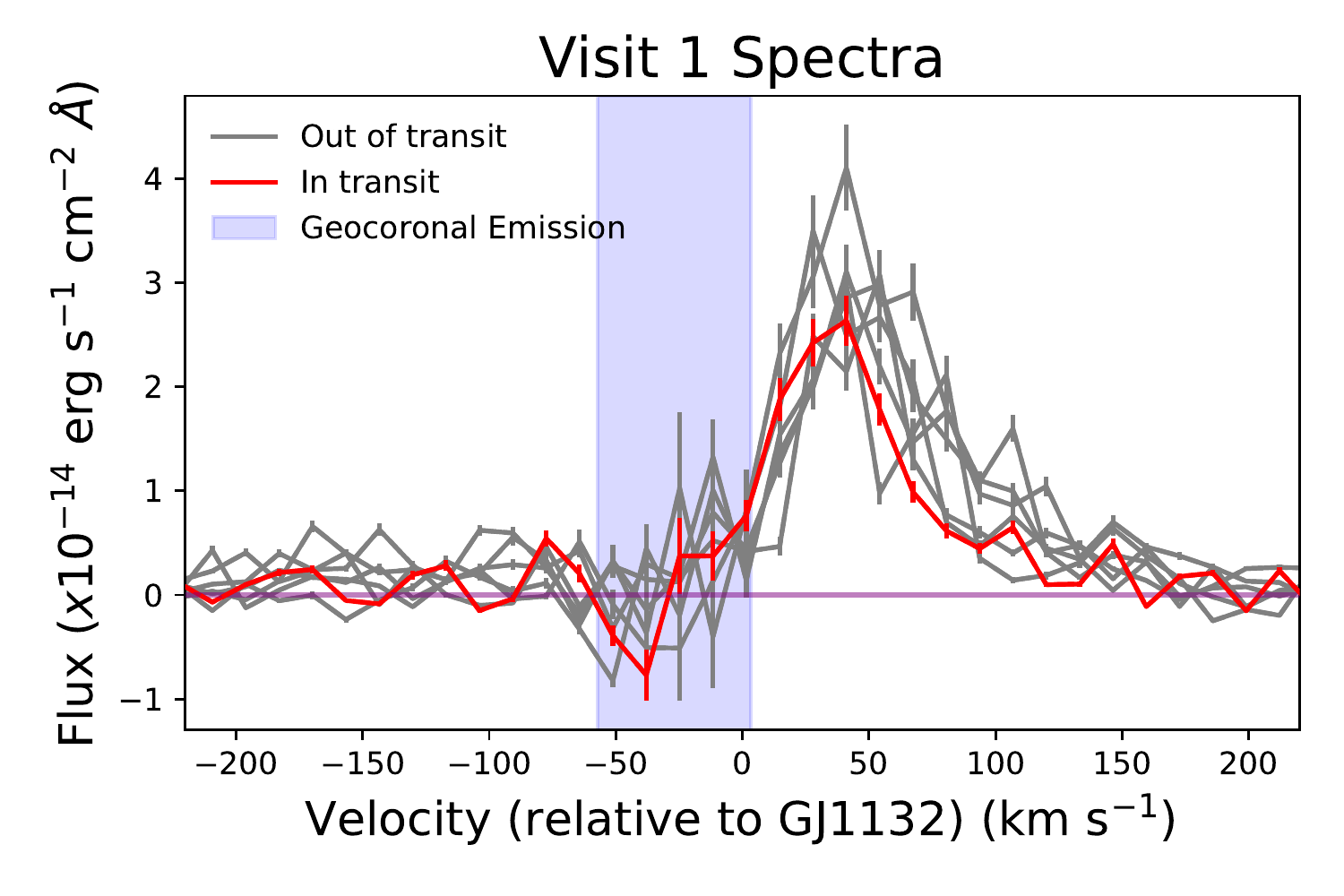}}
\subfloat[]{\includegraphics[width=0.49\textwidth]{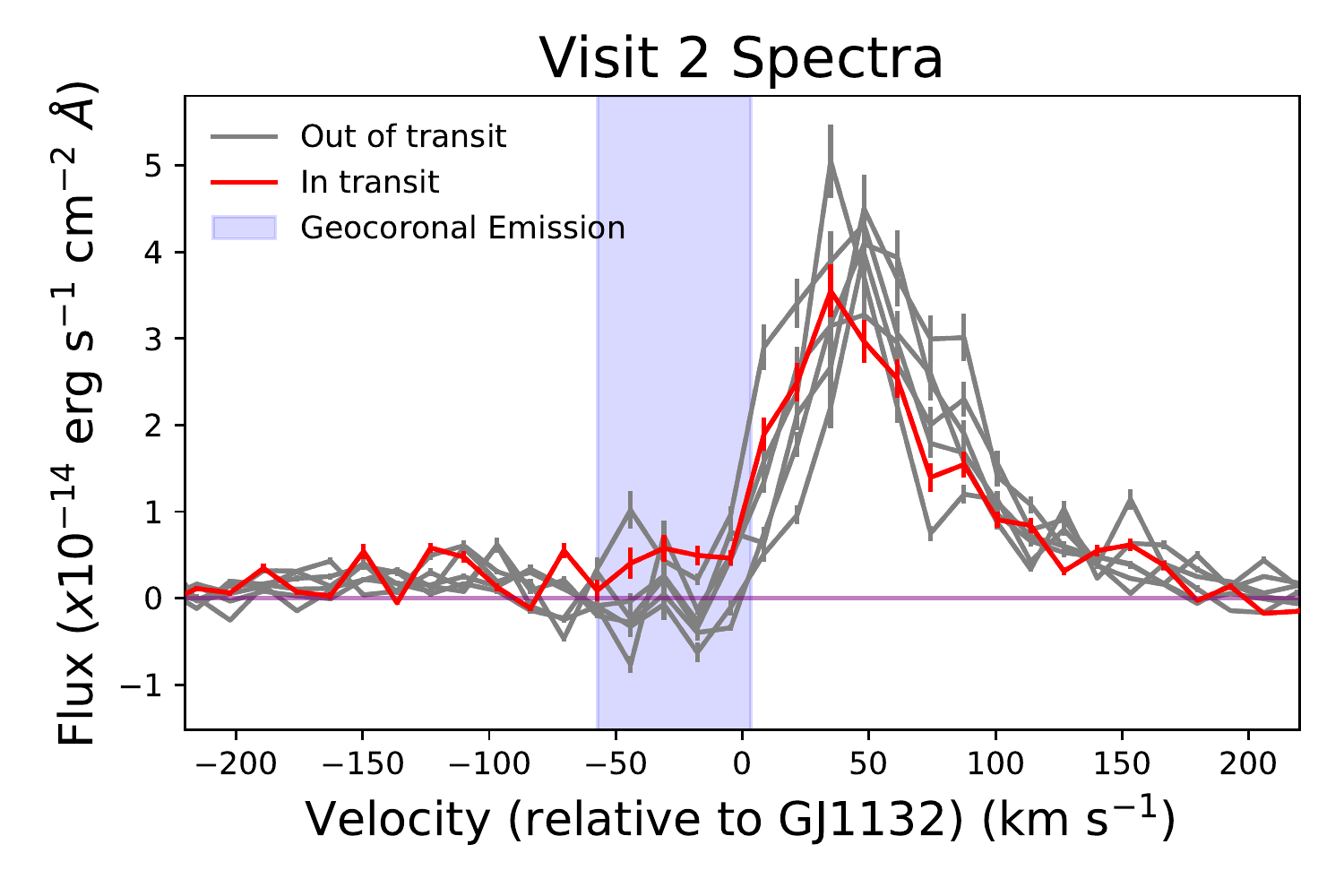}}\\
\subfloat[]{\includegraphics[width=0.59\textwidth]{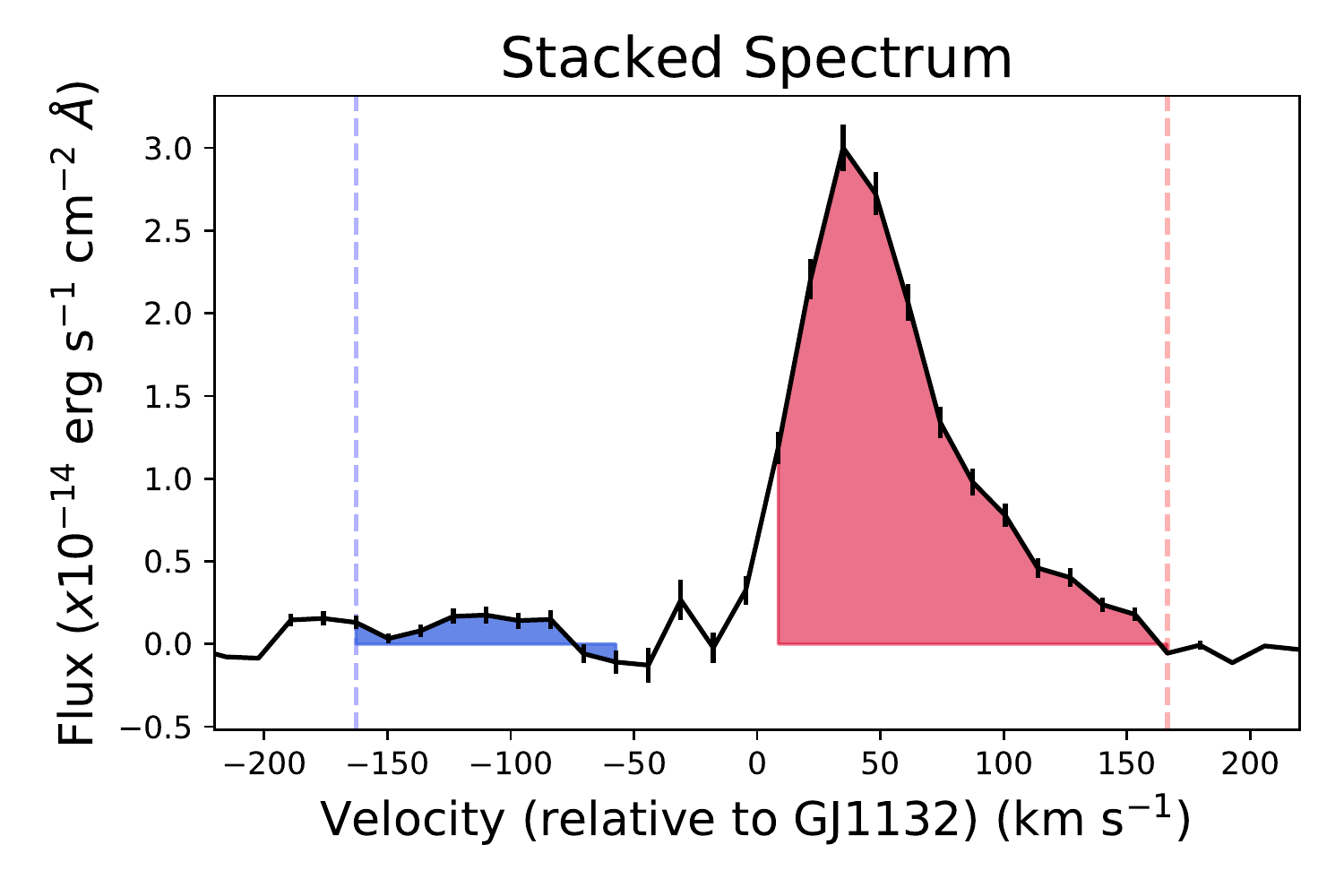}}
\caption{All 14 STIS Ly$\alpha$ spectra in visits 1 (a) and 2 (b) and the averaged stacked spectrum (c). The shape of the stellar Ly$\alpha$ line is a Voigt profile which has been reshaped by convolution with the STIS line spread function and ISM absorption by neutral atomic hydrogen and deuterium. The integration regions for summing up the total Ly$\alpha$ flux are the shaded blue and red areas in (b), with a region in the middle that we omit due to the geocoronal emission. It is apparent that the blue-shifted region of the spectrum is at the noise level, and therefore unlikely to give us any viable information. We set the reference velocity for the spectral profiles at $35~\rm{km~s^{-1}}$, as this is the cited system velocity \citep{Berta-Thompson2015}. \label{Spectra}}
\vspace{0.75cm}
\end{figure*}

\citet{Miguel2015} find that the source of this outflowing hydrogen is from the H$_2$-dominated atmosphere of Gl 436b, with reactions fueled by OH$^-$. Ly$\alpha$ photons from the M dwarf host star dissociate atmospheric H$_2$O into OH and H, which destroy H$_2$. HI at high altitudes where escape is occurring is formed primarily through dissociation of H$_2$ with contributions from the photolyzed H$_2$O.

Modeling of Gl 436b \citep{Bourrier2015,Bourrier2016} demonstrates that the combination of low radiation pressure, low photo-ionization, and charge-exchange with the stellar wind can determine the structure of the outflowing hydrogen, which manifests as a difference in whether the light curve shows a transit in the blue-shifted region of Ly$\alpha$ or the red-shifted region and imprints a specific spectro-temporal signature to the blue-shifted absorption. \citet{Lavie2017} used new observations to confirm the \citet{Bourrier2016} predictive simulations that this exosphere is shaped by charge-exchange and radiative braking.

As giant hydrogen clouds have thus been detected around warm Neptunes \citep[see also the case of GJ 3470b;][]{2018A&A...620A.147B}, it opens the possibility for the atmospheric characterization of smaller, terrestrial planets. \citet{Miguel2015} also find that photolysis of H$_2$O also increases CO$_2$ concentrations. For Earth-like planets orbiting M dwarfs, understanding the photochemical interaction of Ly$\alpha$ photons with water is very important for the evolution and habitability of a planet's atmosphere. 

\subsection{GJ 1132b}

GJ 1132b is a small terrestrial planet discovered through the MEarth project \citep{Berta-Thompson2015}. It orbits a 0.181 M$_\sun$ M dwarf located 12 parsecs away with an orbital period of 1.6 days \citep{Dittmann2017}. Table \ref{tab:system} summarizes its basic properties. This is one of the nearest known transiting rocky exoplanets and therefore provides us with a unique opportunity to study terrestrial atmospheric evolution and composition.

While GJ 1132b is too hot to have liquid surface water, it is important to establish whether this planet and others like it retain substantial atmospheres under the intense UV irradiation of their M dwarf host stars. Knowing whether warm super-Earths such as GJ 1132b regularly retain volatiles such as water in their atmospheres constrains parameter space for our understanding of atmospheric survivability and habitability.

\citet{Diamond-Lowe2018} rule out a low mean-molecular weight atmosphere for this planet by analyzing ground-based transmission spectra at 700-1040~nm. By fitting transmission models for atmospheric pressures of 1-1000~mbar and varying atmospheric composition, they find that all low mean-molecular weight atmospheres are a poor fit to the data, which is better described as a flat transmission spectrum that could be due to a $>$10x solar metallicity or $>$10$\%$ water abundance. Whether these results imply GJ 1132b has a high mean molecular weight atmosphere or no atmosphere at all remains to be seen. If we detect a Ly$\alpha$ transit then this implies UV photolysis of H$_2$O into neutral H and O, leading to outflowing neutral H. The oxygen could recombine into O$_2$ and O$_3$, resulting in a high mean-molecular weight atmosphere, and wholesale oxidation of the surface. 

This work serves as the first characterization of whether there is a neutral hydrogen envelope outflowing from GJ 1132b as well as an opportunity to characterize the deepest (longest integration) Ly$\alpha$ spectrum of any quiet M dwarf of this mass.

\begin{figure*}[t!]
\centering
\includegraphics[width=\textwidth]{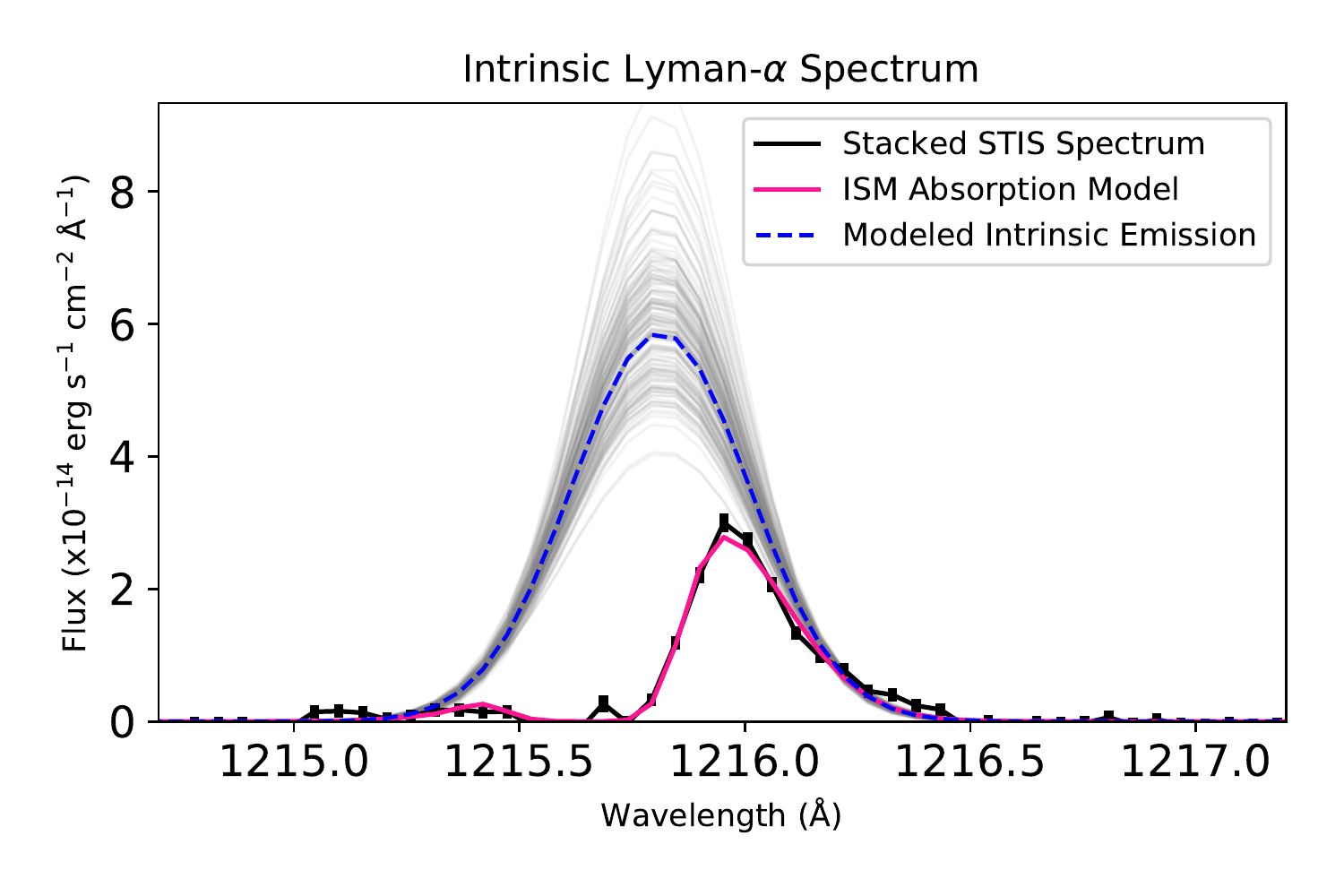}
\caption{Intrinsic Ly$\alpha$ profile for GJ 1132b, with 200 random MCMC samples in gray. The absorption and intrinsic emission models were modeled with the {\tt Lyapy} software which assumes a Voigt profile for the emission and parameterizes the ISM absorption into velocity, line width, and column density. Here, the line center is in the system's rest frame. \label{profile}}
\vspace{0.75cm}
\end{figure*}

\subsection{Solar System Analogs}

The atmospheric evolution and photochemistry we evaluate here is similar to what we have seen in Mars and Venus. Much of Mars' volatile history has been studied in the context of Ly$\alpha$ observations of a neutral H corona that surrounds present-day Mars. \citet{Chaffin2015} use Ly$\alpha$ observations to constrain Martian neutral H loss coronal structure, similar to what we attempt in this work. Indeed, Mars has historically lost H$_2$O via photochemical destruction and escape of neutral H \citep{Nair1994,Zahnle2008}, though the solar wind-driven escape mechanisms for Mars are not necessarily the same as what we propose for GJ 1132b in this work.

Venus has long been the example for what happens when a terrestrial planet is irradiated beyond the point of habitability, as is more than likely the case with GJ 1132b. Venus experienced a runaway greenhouse effect which caused volatile loss and destruction of H$_2$O. \citet{Kasting1983} study the effects of solar UV radiation on an early Venus atmosphere. They find that within a billion years, Venus could have lost most of a terrestrial ocean of water through hydrodynamic escape of neutral H, after photochemical destruction of H$_2$O. GJ 1132b has a higher surface gravity than Venus, which would extend this time scale of hydrogen loss, but it also has a much higher insolation which would reduce the hydrogen loss timescale. Later in this work, we will estimate the expected maximum mass loss rate for GJ 1132b based on the stellar Ly$\alpha$ profile.

The rest of the paper will be as follows. In \S 2 we describe the methods of analyzing the STIS data, reconstructing the stellar spectrum, and analyzing the light curves. In \S 3 we describe the transit fit and intrinsic spectrum results. We discuss the results and their implications in \S 4, including estimates of the mass loss rate from this planet's atmosphere. In \S 5 we describe what pictures of GJ 1132b's atmosphere we are left with.

\begin{figure*}[t!]
\centering
\includegraphics[width=\textwidth]{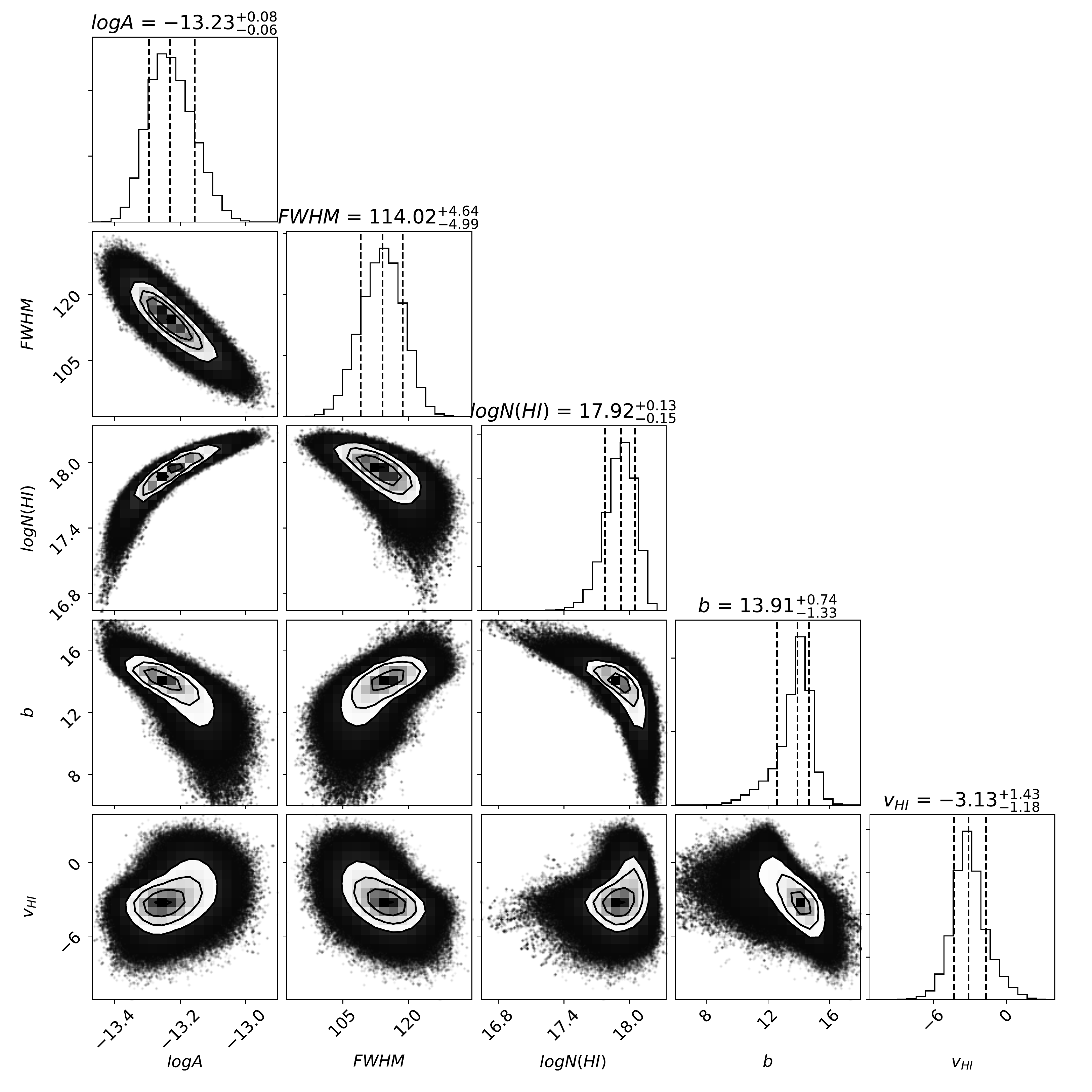}
\caption{Corner plot showing the samples used in recreating the intrinsic emission profile. We omitted the stellar radial velocity samples because the prior was well constrained by independent radial velocity measurements. In this plot, log(A) is the log of the emission amplitude (which has units of erg s$^{-1}$ cm$^{-2}$ \AA$^{-1}$), FWHM is the emission Full Width Half Maximum in km s$^{-1}$, logN(HI) is the log of the column density of neutral ISM hydrogen (which has units of cm$^{-2}$), b is the ISM Doppler parameter in km s$^{-1}$, and v$_{\rm{HI}}$ is the ISM cloud velocity in km s$^{-1}$. \label{corner}}
\vspace{0.75cm}
\end{figure*}
   
\section{Methods}

\subsection{Hubble STIS Observations}

To study the potential existence of a neutral hydrogen envelope around this planet, we scheduled 2 transit observations of 7 orbits each (2 observations several hours from mid-transit for an out of transit measurement and 5 observations spanning the transit) with the Space Telescope Imaging Spectrograph (STIS) on the Hubble Space Telescope (HST)\footnote{Cycle 24 GO proposal 14757, PI: Z Berta-Thompson}. We used the G140M grating with the 52”x0.05” slit, collecting data in TIME-TAG mode with the FUV-MAMA photon-counting detector. This resulted in 14 spectra containing the Ly$\alpha$ emission line (1216~\AA), which show a broad profile that has been centrally absorbed by neutral ISM atomic hydrogen.

We re-extracted the spectra and corrected for geocoronal emission using the {\tt calstis} pipeline \citep{Hodge1995}. The STIS spectrum extraction involved background subtraction which accounts for geocoronal emission (see Fig. \ref{x2d}), leaving us only with the need to model the stellar emission and ISM absorption. We omit data points from both visits that fall within the geocoronal emission signal, wavelengths from both visits that overlapped with strong geocoronal emission and therefore had high photon noise. We thus define our blue-shifted region to be $<$-60~km~s$^{-1}$ and our red-shifted region to be $>$10~km~s$^{-1}$ relative to the star. One potential source of variability is where the target star falls on the slit. If it fell directly on the slit, then the observed flux will be more than if the star was partially off the slit. To account for this, we scheduled ACQ/PEAK observations at the start of each HST orbit to center the star on the slit and minimize this variability.

In order to analyze the light curves with higher temporal resolution, we used the STIS time-tag mode to split each of the 14 2~ks exposures into 4 separate 0.5~ks sub-exposures. This detector records the arrival time of every single photon, which is what allows us to create sub-exposures in time-tag mode. Each 2D spectrum sub-exposure was then converted into a 1D spectrum. To do this, we first defined an extraction window around the target spectrum (see Fig. \ref{x2d}) and summed up all the flux in that window along the spatial axis. Extraction windows were also defined on either side of the target in order to estimate the background and subtract that from the target window. This results in a noisy line core but eliminates the geocoronal emission signature (Fig. \ref{Spectra}a \& \ref{Spectra}b). These steps were all performed with {\tt calstis}.

\begin{figure*}[t!]
\centering
\subfloat[]{\includegraphics[width=0.49\textwidth]{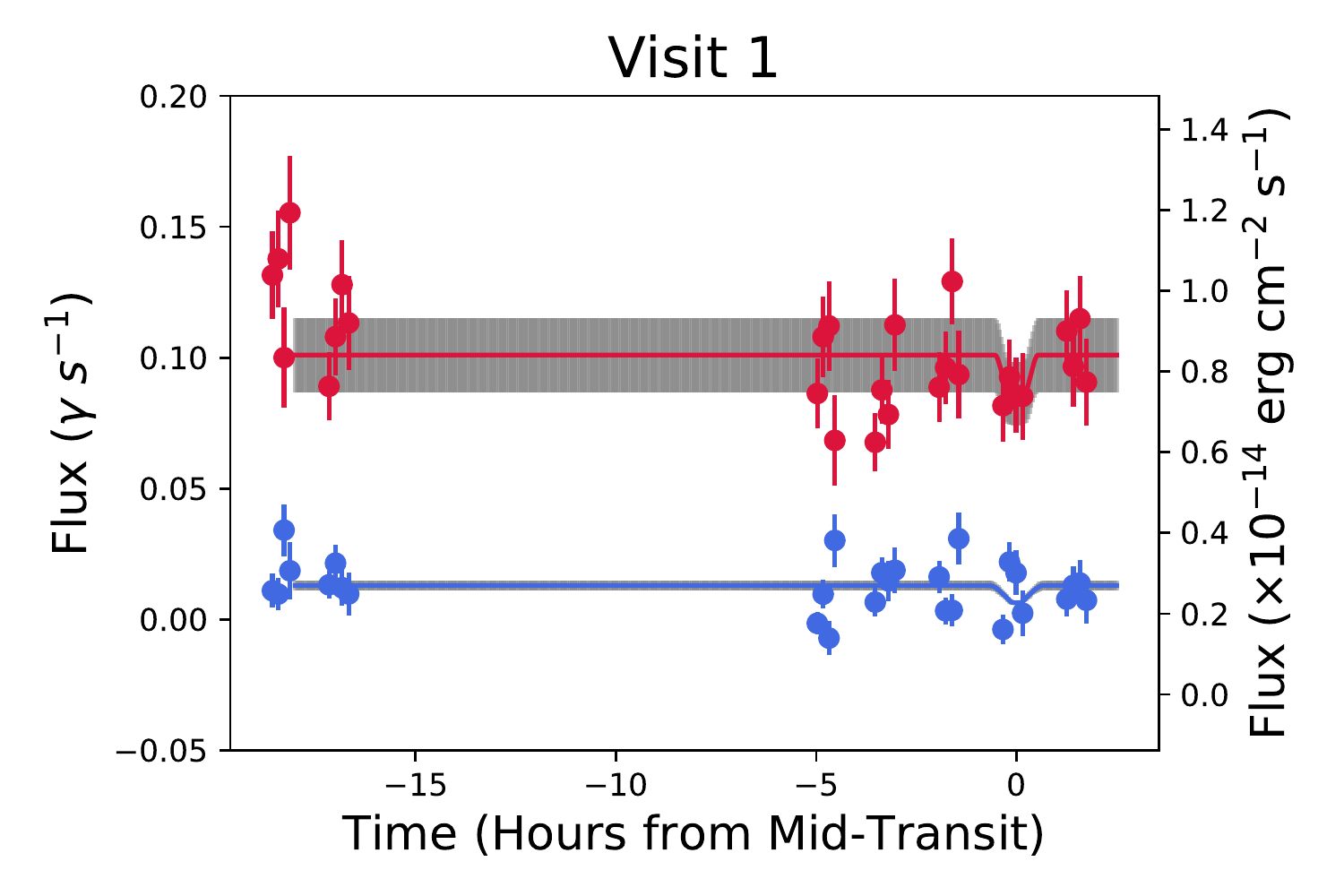}}
\subfloat[]{\includegraphics[width=0.49\textwidth]{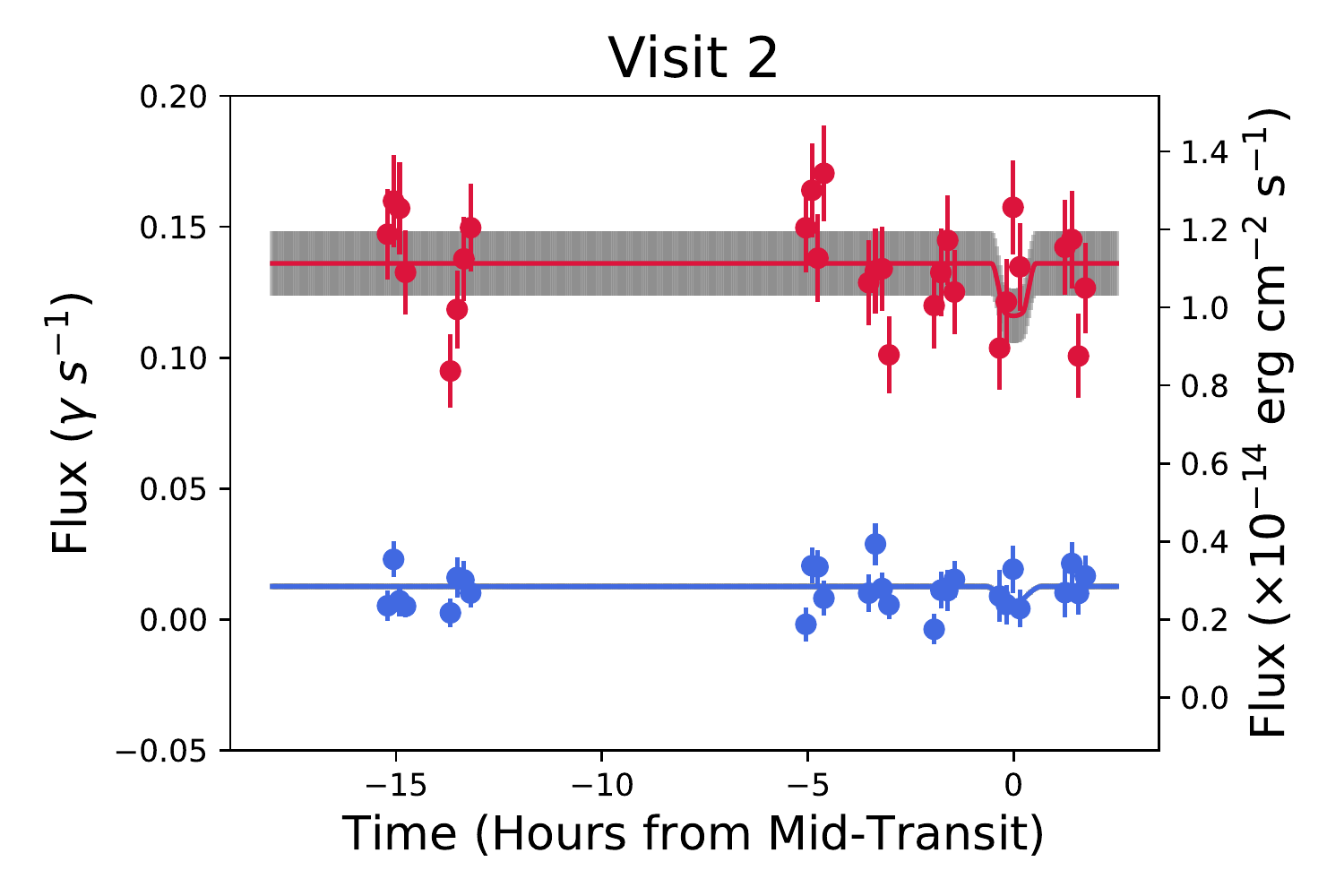}}
\caption{Modeled light curves from both visits. In addition to the calibrated flux values, we display the flux in photons~s$^{-1}$ because the SNR is very low at Ly$\alpha$ and this motivated us to use a Poisson likelihood in our analysis of the light curves. Some data points fall to negative values, which can happen when the data point has effectively no flux and then data reduction processes (such as background subtraction) subtract a slightly higher amount of flux. The gray bars indicate what we calculate as a 15\% "stellar variability" fudge factor - acquired by calculating what size of error bars would be necessary to result in a $\chi^2$ value of 1 for our best fit models. The blue wing light curves don't provide much information due to their extremely low flux but we can see from the red wing fits that there is an upper limit on the transit depth.\label{lightcurves}}
\vspace{0.75cm}
\end{figure*}

\subsection{Stellar Spectrum Reconstruction}

With the same spectra used for light curve analysis, we created a single weighted average spectrum, representing 29.3~ks (8.1~hrs) of integration at Ly$\alpha$ across 14 exposures (Fig. \ref{Spectra}c). This stacked spectrum was used with {\tt LyaPy} modeling program \citep{Youngblood2016} that uses a 9-dimensional MCMC to reconstruct the intrinsic stellar spectrum assuming a Voigt profile. Modeling observed Ly$\alpha$ spectra is tricky because of the neutral ISM hydrogen found between us and GJ 1132. This ISM hydrogen has its own column density, velocity, and line width which creates a characteristic absorption profile within our Ly$\alpha$ emission line.

This model takes 3 ISM absorption parameters (column density, cloud velocity, Doppler parameter) and models the line core absorption while simultaneously modeling the intrinsic emission which would give us the resulting observations. Turbulent velocity of the ISM is assumed to be negligible, with the line width dominated by thermal broadening. A fixed deuterium-to-hydrogen ratio of $1.56\times10^{-5}$ \citep{Wood2004} is also applied to account for the deuterium absorption and emission near Ly$\alpha$. Modeling the ISM parameters required us to approximate the local interstellar medium as a single cloud with uniform velocity, column density, and Doppler parameter. While the local ISM is more complex than this single component and contains two clouds (G, Cet) in the line of sight toward GJ~1132 \citep[based on the model described in][]{Redfield2000}, our MCMC results strongly favored the velocity of the G cloud, so we defined the ISM priors based on this cloud \citep{Redfield2000,Redfield2008}.

We use uniform priors for the emission amplitude and FWHM, and Gaussian priors for the HI column density, stellar velocity, HI Doppler width, and HI ISM velocity. The HI column density and Doppler width parameter spaces were both truncated in order to prevent the model from exploring physically unrealistic values. For N$_{\rm{HI}}$, we restrict the parameter space to 10$^{16}$-10$^{20}$ cm$^{-2}$, based on the stellar distance (12.04~pc) and typical n$_{\rm{HI}}$ values of $0.01-0.1$~cm$^{-3}$ \citep{Redfield2000,Wood2005}. We limit the Doppler width to 6-18 km~s$^{-1}$, based on estimates of the Local Interstellar Cloud (LIC) ISM temperatures \citep{Redfield2000}.

\subsection{Light Curve Analysis}

The extracted 1D spectra were then split into a blue-shifted regime and red-shifted regime, on either side of the Ly$\alpha$ core (Fig. \ref{Spectra}c) so that we could integrate the total blue-shifted and red-shifted flux and create 4 total light curves from the 2 visits (Fig. \ref{lightcurves}). Each of these light curves was fitted with a {\tt BATMAN} \citep{Kreidberg2015} light curve using a 2-parameter MCMC with the {\tt emcee} package \citep{Foreman-Mackey2012}. The {\tt BATMAN} models assume that the transiting object is an opaque disk, which is usually appropriate for modeling planetary sizes. However, we are modeling a possible hydrogen exosphere which may or may not be disk-like, and which would have varying opacity with radius. For this work, we use the {\tt BATMAN} modeling software with the understanding that our results tell us the effective radius of a cartoon hydrogen exosphere, with an assumed spherical geometry.

\begin{figure*}[t!]
\includegraphics[width=\textwidth]{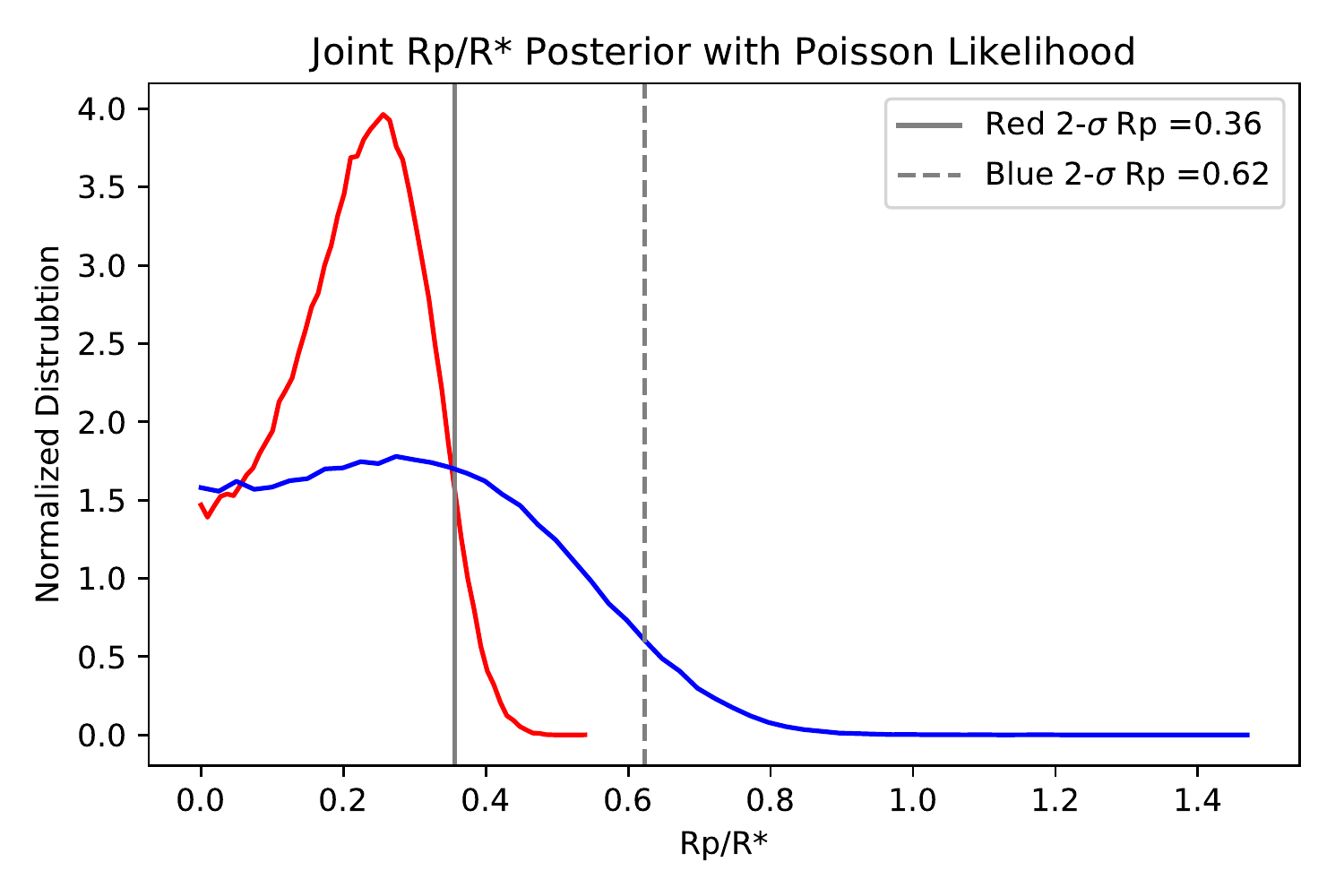}
\caption{Joint posterior distribution for the R$_p$/R$_*$ distributions for both visits. Poisson likelihoods were used due to the low photon count regime of these spectra.\label{2sig}}
\vspace{0.75cm}
\end{figure*}

We fit for R$_p$/R$_*$ and the baseline flux using a Poisson likelihood for each visit. We use a Poisson distribution because at Ly$\alpha$, the STIS detector is receiving very few photons. Our log(likelihood) function is:
$$\ln(likelihood)=\sum_i[d_i\ln(m_i)-m_i-\ln(d_i!)]$$ where $d_i$ is the total (\textit{gross}) number of photons detected and $m_i$ is the  modeled number of photons detected. The photon model is acquired by taking a {\tt BATMAN} model of in-transit photons and adding the \textit{sky} photons, which is data provided through the {\tt calstis} reduction pipeline. Uniform priors are assumed for both R$_p$/R$_*$ and the baseline flux. We restrict our parameter space to explore only effective cloud radii $>$~0, representing physically plausible clouds that block light during transit. By taking simple averages of the light curve fluxes, we find the ratio of the in-transit flux compared with out-of-transit flux to be $1.01\pm0.16$ for the visit 1 red-wing flux and $0.97\pm0.13$ for the visit 2 red-wing. As both are consistent with no detectable transit, the constraints we obtain from the fitting procedure will represent upper limits on the effective size of any hypothetical cloud.

\section{Results}

\subsection{Spectrum Reconstruction}

Figure \ref{profile} shows the best fit emission model with 1-sigma models and a corner plot to display the most crucial modeling parameters, with MCMC results shown in Table \ref{tab:intrinsic} and Figure \ref{corner}. This result gives us the total Ly$\alpha$ flux for this M dwarf.

The results of the stellar spectrum reconstruction indicate that there is one component of Ly$\alpha$ flux, though that is potentially a result of the low SNR regime of these observations. Additionally, our fit indicates that there is one dominant source of ISM absorption between us and GJ 1132 - a single cloud with velocity $-3.1$~km~s$^{-1}$, HI column density $10^{17.9}$~cm$^{-2}$ and Doppler parameter $13.9$ km~s$^{-1}$. Our current understanding of LIC \citep{Redfield2000,Redfield2008} indicates that there should be 2 clouds, \textit{G} and \textit{Cet} in the line of sight of GJ 1132, but our derived v$_{HI}$ is consistent with the velocity of \textit{G}, which is reported as $-2.73\pm0.94$~km~s$^{-1}$. We take this to mean that the \textit{G} cloud is the dominant source of absorption and that we can subsequently reconstruct this spectrum under a single-cloud assumption.

\begin{figure*}[t!]
\includegraphics[width=\textwidth]{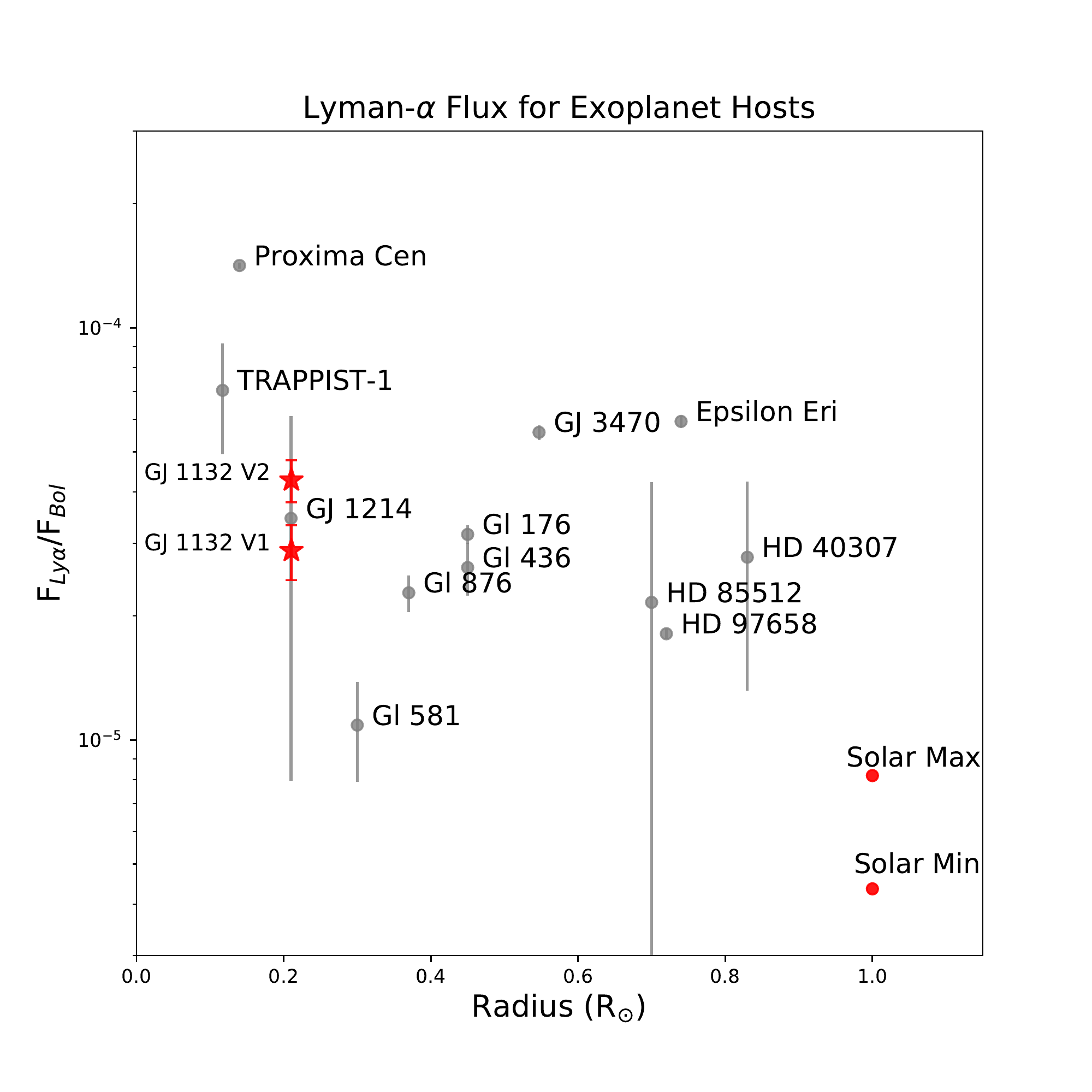}
\caption{Comparison of F[Ly$\alpha$]/F[bol] for GJ 1132 compared with stars in the MUSCLES Treasury Survey \citep{Youngblood2016,Youngblood2017}, TRAPPIST-1 \citep{Bourrier2017}, HD 97658 \citep{Bourrier2017a}, GJ 436 \citep{Bourrier2015}, GJ 3470 \citep{Bourrier2018}, as well as the Sun \citep{Linsky2013}. The stars shown here are all M and K dwarfs that are known exoplanet hosts. The error bars on GJ 1132 are statistical errors based on our modeling, so we have included the flux ratios from both visits (9 months apart) to display the variability we see in the data, labeled V1 and V2.\label{MUSCLES}}
\vspace{0.75cm}
\end{figure*}

By integrating the reconstructed emission profile, we find a Ly$\alpha$ flux of 2.88$^{+0.42}_{-0.31}$x10$^{-14}$ erg~s$^{-1}$~cm$^{-2}$ which gives f[Ly$\alpha$]/f[bol] = 2.9$\pm$0.4x10$^{-5}$, where we have calculated the bolometric luminosity of GJ 1132b as:
\begin{equation}
    f_{\rm{bol}} = \sigma T_{\rm{eff}}^4\left(\frac{R_*}{\rm{distance}}\right)^2,
\end{equation}
Where values for the T$_{\rm{eff}}$ and R$_*$ were taken from \citet{Bonfils2018} and the distance to the star is taken from \citet{Dittmann2017}.
Compared with the Sun which has f[Ly$\alpha$]/f[bol] = 4.6x10$^{-6}$ \citep{Linsky2013}, we can see that this M dwarf emits fractionally 6x more of its radiation in the ultraviolet.

Given the intra-visit stellar variability, we also modeled the average Ly$\alpha$ spectra for visits 1 and 2 separately. All modeled parameters (see Fig. \ref{corner}) were consistent between visits except the FWHM, which were different by 3-$\sigma$, and the total integrated fluxes which differed by 2-$\sigma$ (2.90$^{+0.47}_{-0.41}$x10$^{-14}$ erg~s$^{-1}$~cm$^{-2}$ for visit 1 and 4.30$^{+0.52}_{-0.43}$x10$^{-14}$ erg~s$^{-1}$~cm$^{-2}$ for visit 2). For the calculation of mass loss rates in section \S 4.1, we use the integrated flux of the combined reconstructed spectrum (Fig. \ref{profile}).

\subsection{Light Curve Modeling}

The light curves for both visits are shown in Figure \ref{lightcurves}. MCMC modeling of these light curves resulted in best fit parameters shown in Table \ref{tab:curve}. We report no statistically significant transits, but we can use the modeling results to calculate limits on the hydrogen cloud parameters. To ensure that we were not biasing our results by converting from the measured flux counts to photons~s$^{-1}$, we also analyzed the flux-calibrated light curves with Gaussian likelihoods based on pipeline errors and found the results did not significantly differ from what we present here.

\subsubsection{The STIS Breathing Effect}

There is a well-known intra-orbit systematic which shows up in Hubble STIS observations known as the \textit{breathing effect} which can result in a change of amplitude of about $0.1\%$ over the course of an HST orbit. \citep[e.g.,][]{2001ApJ...552..699B,2008ApJ...686..658S,Bourrier2017}. This effect is small compared to the photon uncertainty in these observations, but to examine this STIS systematic, we perform our light curve analysis on the non-time-tagged data. We find that the results are consistent with our time-tagged analysis, so we posit that this effect does not significantly alter our conclusions.

\subsection{Stellar Variability}

The red wing of our spectral data show a highly variable stellar Ly$\alpha$ flux over the course of these HST visits and we quantify this variability as a Gaussian uncertainty,
\begin{equation}
    \sigma_x^2 =\sigma_{\rm{measured}}^2-\sigma_{\rm{photometric}}^2,
\end{equation}
where $\sigma_{\rm{measured}}$ is our RMS noise and $\sigma_{\rm{photometric}}$ is the \texttt{calstis}-generated error propagated through our spectral integration. Within one 90-minute HST orbit, we see flux variabilities ($\sigma_x$) of 5-16\% for visit 1 and 7-18\% for visit 2. Among one entire 18-hour visit, variability is 20\% for visit 1 and 14\% for visit 2 while in the 9 months between the two visits, there is a 22\% offset. These results are consistent with the 1-41\% M dwarf UV variability found by \citet{Loyd2014}.

\section{Discussion}

With 14 STIS exposures, we have characterized a long-integration Ly$\alpha$ spectrum and furthered our understanding of the intensity of UV flux from this M dwarf. \citet{France2012} find that as much as half of the UV flux of quiescent M dwarfs is emitted at Ly$\alpha$, so knowing the total amount of flux at this wavelength serves as a proxy for the total amount of UV flux for this type of star. Our measurement of this Ly$\alpha$ flux provides a useful input for photochemical models of haze, atmospheric escape, and molecular abundances in this planet's atmosphere. 

From the red-shifted light curves, we can calculate a 2-$\sigma$ upper limit on the radius of this potential hydrogen cloud outflowing from GJ 1132b. We calculate this upper limit (see Fig. \ref{2sig}) by taking the joint (visit 1 \& visit 2) posterior distributions that resulted from MCMC modeling of these light curves and integrating the CDF to the 95\% confidence interval and examining the corresponding R$_p$/R$_*$. The 2-$\sigma$ upper limit from the red-shifted Ly$\alpha$ spectra gives us an R$_p$/R$_*$ of 0.36. The upper limit R$_p$/R$_*$ from the blue-shifted light curves is 0.62 but given the very low SNR of that data, this is not a meaningful constraint. The red-shifted result is an upper limit on the effective radius of a hydrogen coma, and the real coma could be much more diffuse and asymmetric.

\begin{table}
\begin{center}
\begin{tabular}{cc}
\toprule
    Line Velocity [km~s$^{-1}$] &  35.23$^{+0.99}_{-0.98}$\\
    \midrule
    log(Amplitude) [erg~s$^{-1}$~cm$^{-2}$~{\rm{\AA}$^{-1}$}] & -13.23$^{+0.08}_{-0.06}$ \\
    \midrule
    FWHM [km~s$^{-1}$] &  114.02$^{+4.64}_{-4.99}$\\
    \midrule
    log(HI Column Density) [cm$^{-2}$] &  17.92$^{+0.13}_{-0.15}$\\
    \midrule
    Doppler Parameter (b) [km~s$^{-1}$] &  13.91$^{+0.74}_{-1.33}$\\
    \midrule
    HI Velocity [km~s$^{-1}$] &  -3.13$^{+1.43}_{-1.18}$\\
    \midrule
    Total Flux [erg~s$^{-1}$~cm$^{-2}$] & ${2.9\times10^{-14}}^{+4\times10^{-15}}_{-3\times10^{-15}}$\\
    \midrule
    Total Flux (1 Au) [erg~s$^{-1}$~cm$^{-2}$] & ${0.18}^{+0.03}_{-0.02}$\\
\bottomrule
\end{tabular}
\end{center}
\caption{Intrinsic emission line model parameters taken from MCMC samples, with 1-$\sigma$ error bars. \textit{Total Flux (1 Au)} is the flux if it were measured 1 Au from the star, whereas the \textit{Total Flux} is the flux as measured at HST.\label{tab:intrinsic}}
\vspace{0.2cm}
\end{table}

\subsection{GJ 1132b Atmospheric Loss}
In order to connect our results to an upper limit on the possible mass loss rate of neutral H from this planet's atmosphere, we follow the procedure outlined in \citet{Kulow2014}.

Assuming a spherically symmetric outflowing cloud of neutral H, the equation for mass loss is \begin{equation}
    \dot{M}_{HI} = 4\pi r^2v(r)n_{HI}(r)
\end{equation}
Where $v(r)$ is the outflowing particle velocity and $n_{HI}(r)$ is the number density of HI at a given radius, r. For this calculation, we will be examining our 2-$\sigma$ upper limit radius at which the cloud becomes optically thick, where (R$_p$/R$_*$)$^2$ = $\delta$ = 0.13. We assume a $v$ range of $10-100$~km~s$^{-1}$, which is the range of the planet's escape velocity ($10$~km~s$^{-1}$) and the stellar escape velocity ($100$~km~s$^{-1}$).

\citet{Kulow2014} reduce Equation (3) to
\begin{equation}
    \dot{M}_{HI} = \frac{2\delta{R_*}mv}{\sigma_0}
\end{equation}
with a Ly$\alpha$ absorption cross-section $\sigma_0$ defined as
\begin{equation}
    \sigma_0 = \frac{\sqrt{\pi}e^2}{m_ec\Delta\nu_D}f
\end{equation}
where $e$ is the electron charge, $m_e$ is the electron mass, $c$ is the speed of light, $f$ is the particle oscillator strength (taken to be $0.4161$ for HI) and $\Delta\nu_D$ is the Doppler width, $b/\lambda_0$, where we use $100~\rm{km~s^{-1}}$ for b, as was done in \citet{Kulow2014}.

This gives us an upper limit mass loss rate of $\dot{M}_{HI} < 0.86\times10^{9}$~g~s$^{-1}$ for neutral hydrogen, corresponding to $15.4\times10^9$~g~s$^{-1}$ of water decomposition, assuming all escaping neutral H comes from H$_2$O. If this upper-limit mass loss rate was sustained, GJ 1132b  would lose an Earth ocean in approximately $6$~Myr. If we had actually detected mass loss at this high rate, it would likely indicate that there had been recent delivery or outgassing of water on GJ 1132b, because primordial atmospheric water would have been lost on time scales much shorter than the present age of the system.

\begin{table}[b!]
\vspace{0.5cm}
\begin{center}
\begin{tabular}{cccc}
\toprule
    \textbf{MCMC Results} & Visit 1 & Visit 2 & Joint\\
    \midrule
    R$_p$/R$_*$ (R) & 0.34$^{+0.11}_{-0.15}$ & 0.15$^{+0.12}_{-0.10}$ & 0.22$^{+0.09}_{-0.12}$\\
    \midrule
    R$_p$/R$_*$ (B) & 0.29$^{+0.24}_{-0.20}$ & 0.46$^{+0.30}_{-0.30}$ & 0.30$^{+0.21}_{-0.21}$\\
    \midrule
    Baseline ($\gamma$~s$^{-1}$) (R) & 0.102$^{+0.003}_{-0.003}$ & 0.136$^{+0.003}_{-0.004}$ & 0.101$^{+0.003}_{-0.003}$\\
     &  &  & 0.136$^{+0.003}_{-0.003}$\\
    \midrule
    Baseline ($\gamma$~s$^{-1}$) (B) & 0.013$^{+0.001}_{-0.001}$ & 0.013$^{+0.001}_{-0.002}$ & 0.013$^{+0.001}_{-0.001}$\\
    & & & 0.013$^{+0.001}_{-0.002}$\\
\bottomrule
\end{tabular}
\caption{Light curve fit results for MCMC sampling where Poisson likelihoods were used.\label{tab:curve}}
\end{center}
\end{table}

\begin{figure*}
\centering
\subfloat[]{\includegraphics[width=0.49\textwidth]{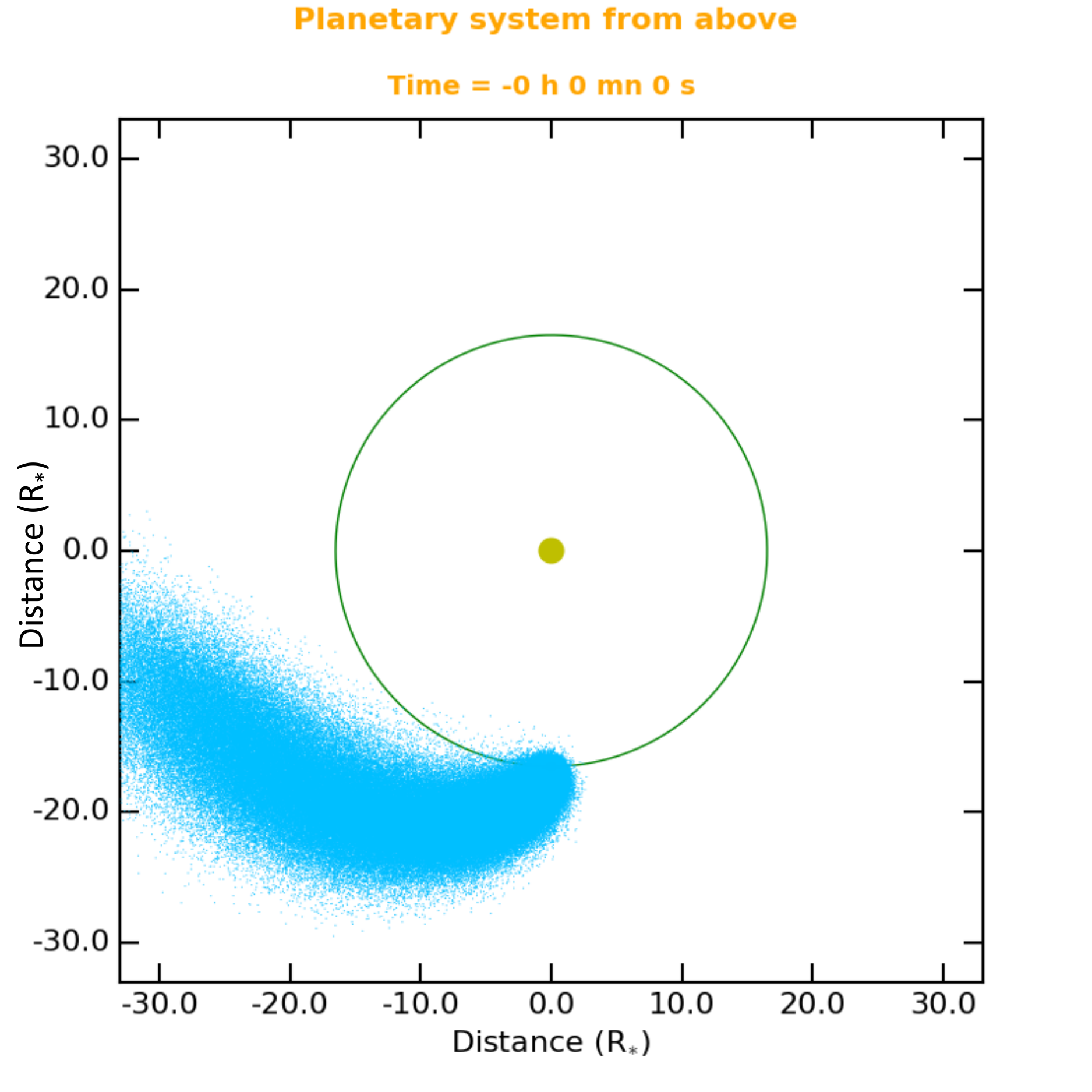}}
\subfloat[]{\includegraphics[width=0.49\textwidth]{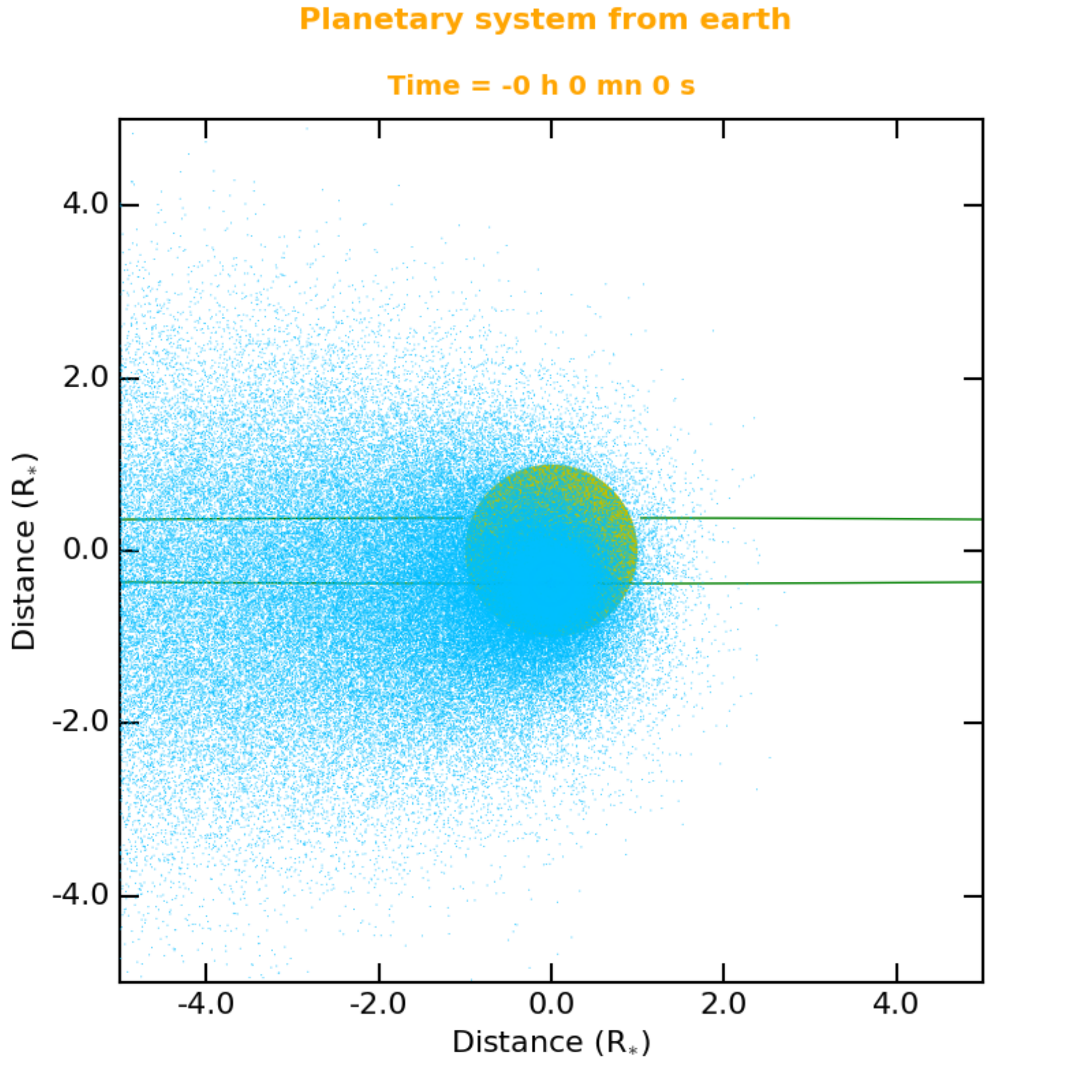}}
\caption{Simulations of the GJ1132 system showing the dynamics of a hypothetical outflowing hydrogen cloud. The left panel shows a top-down view of the system, as a hydrogen tail extends in a trailing orbit. The right panel shows the view from an Earth line of sight, at mid-transit.\label{simulations}}
\vspace{0.75cm}
\end{figure*}

We can also calculate the energy limited mass loss rate, corresponding to the the ratio of the incoming XUV energy to the work required to lift the particles out of the atmosphere:

\begin{equation}
    \dot{M} = \frac{F_{XUV}\pi R_p^2}{GM_p~R_p^{-1}} = \frac{F_{XUV}\pi R_p^3}{GM_p}.
\end{equation}

The total F$_{\rm{XUV}}$ is the flux value at the orbit of GJ 1132b. Using our derived Ly$\alpha$ flux, the {\tt Lyapy} package calculates stellar EUV spectrum and luminosity from 100-1171 \AA~based on \citet{2014ApJ...780...61L}. From that EUV spectrum, we then calculate the 5-100 \AA~XUV flux based on relations described in \citet{King2018}.

Assuming 100$\%$ efficiency, we obtain an energy-limited neutral hydrogen mass loss rate of $3.0\times10^9$~g~s$^{-1}$ estimated from the stellar spectrum reconstruction. This energy-limited escape rate is commensurate with the upper-limit we calculate based on the transit depth and stellar properties in the previous section. If we assume a heating efficiency of $1\%$ \citep[based on similar simulations done in][]{Bourrier2016}, then we arrive at a low expected neutral hydrogen loss rate of $3.0\times10^7$~g~s$^{-1}$, below the level of detectability with these data.

\subsection{Simulating HI Outflow from GJ 1132b }

Figure \ref{simulations} shows simulation results for neutral hydrogen outflowing from GJ 1132b from the EVaporating Exoplanet code ({\tt EVE}) \citep{Bourrier2013, Bourrier2016}. This code performs a 3D numerical particle simulation given stellar input parameters and atmospheric composition assumptions. These simulations were performed using the Ly$\alpha$ spectrum derived in this work, where the full XUV spectrum has been found as described in the previous section. This spectrum is used directly in {\tt EVE} to calculate the photoionization of the neutral H atoms and calculate theoretical Ly$\alpha$ spectra during the transit of the planet as they would be observed with HST/STIS. In addition, our Ly$\alpha$ spectrum is used to calculate the radiation pressure felt by the escaping neutral hydrogen, which informs the dynamics of the expanding cloud.

\begin{figure*}
\includegraphics[width=\textwidth]{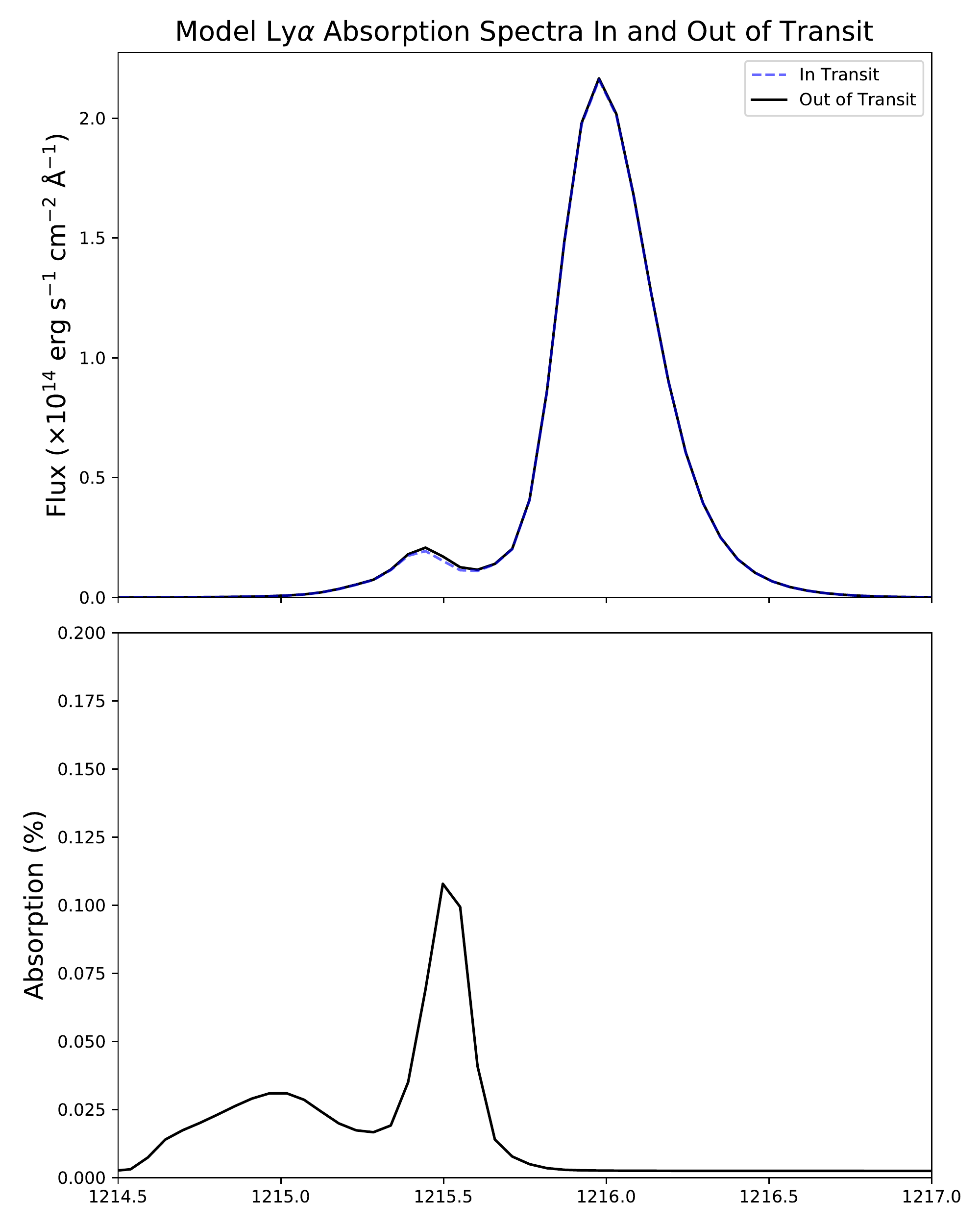}
\caption{{\tt EVE} simulated absorption spectra in-transit and 4 hours pre-transit. We can see that the only region of significant absorption is at $1215.5$~\AA, where absorption peaks at about 12$\%$ as seen in the bottom panel. While there is a larger expected flux decrease in the blue wing, the signal is largely in the region that the ISM absorbs and our data are too noisy in the blue wing to detect the possible absorption signal seen in the models. The mass loss rate corresponding to the above model is 1$\times$10$^7$g~s$^{-1}$.\label{modelspectra}}
\end{figure*}

{\tt EVE} simulations were created with the following assumptions: The outflowing neutral hydrogen atoms escape from the Roche lobe altitude ($\sim5~R_p$) at a rate of $1\times10^7$ g~s$^{-1}$, modeled as a Maxwellian velocity distribution with upward bulk velocity of 5~km~s$^{-1}$ and temperature of 7000~K, resulting in a cloud which could absorb upwards of 80$\%$ of the flux in the blue wing. However, GJ 1132 has a positive radial velocity, so blue-shifted flux falls into the regime of ISM absorption and the signal is lost. Simulations of the in-transit and out-of-transit absorption spectra as they would be observed at infinite resolution by HST are shown in Figure \ref{modelspectra}. However, the simulations don't rule out that some thermospheric neutral H may absorb some extra flux in the red wing \citep[see][for a justification of simulation parameters]{Salz2016}. We note that for planets around M dwarfs, the upward velocity may have a strong influence on the extension of the hydrogen coma. The thermosphere is simulated as a 3D grid within the Roche Lobe, defined by a hydrostatic density profile, and the temperature and upward velocity from above. The exosphere is collisionless with its dynamics dominated by radiation pressure.

There might be other processes shaping the exosphere of GJ 1132b (magnetic field, collisions with the stellar wind, the escaping outflow remaining collisional at larger altitudes than the Roche lobe), but for these simulations we take the simplest possible approach based on what we actually know of the system. Finally, we do not include self-shielding effects of HI atoms within the exosphere, as we do not expect the exosphere is dense enough for self-shielding to significantly alter the results.

The integrated Ly$\alpha$ spectrum corresponds with a maximum ratio of stellar radiation pressure to stellar gravity of 0.4, which puts this system in the regime of radiative breaking \citep{Bourrier2015}, which has a slight effect of pushing neutral hydrogen to a larger orbit. However, the gas is not blown away so the size of the hydrogen cloud will increase if we increase the outward particle velocity. Since the exosphere is not accelerated, most of its absorption is close to 0~km~s$^{-1}$ in the stellar reference frame, with some blue-shifted absorption because atoms in the tail move to a slightly larger orbit than the planet. This indicates that the lack of blue-shifted flux in our observations, due to ISM absorption, is a hindrance to fully understanding the possible hydrogen cloud around this planet. The upper limit cloud size that we quote is based on the observed red-shifted flux in a system which is moving away from us at 35~km~s$^{-1}$, so any cloud absorption of flux closer to the line center is outside of the scope of what we can detect.

\section{Conclusions}

In this work we make the first characterization of the exosphere of GJ 1132b. Until a telescope like LUVOIR \citep{Roberge2018}, these observations will likely be the deepest possible characterization for Ly$\alpha$ transits of this system. If this planet has a cloud of neutral hydrogen escaping from its upper atmosphere, the effective size of that cloud must be less than $0.36$~R$_*$ ($7.3$~R$_{\rm{p}}$) in the red-shifted wing. The blue wing indicates an upper limit of $0.62$~R$_*$ ($12.6$~R$_{\rm{p}}$), though this is a very weak constraint. In addition, we were able to model the intrinsic Ly$\alpha$ spectrum of this star.

This Ly$\alpha$ transit's upper limit R$_p$/R$_*$ implies a maximum hydrogen escape rate of $0.08-0.8\times10^9$~g~s$^{-1}$. If this is the case, GJ 1132b loses an Earth ocean of water between $6-60$~Myr. Since the mass loss rate scales linearly with $F_{\rm{XUV}}$, we estimate that if this planet were in the habitable zone of its star, about 5x further than its current orbit \citep[based on HZ estimates in][]{Shields2016}, the planet would lose an Earth ocean of water in as little as 0.15-1.5 Gyr. However, these values are based on 2-$\sigma$ upper limits and theoretical calculations suggest mass loss rates lower than these values, so further Ly$\alpha$ observations are needed to better constrain this mass loss. In addition, these estimates are based on the current calculated UV flux of GJ 1132, which likely decreases over the star's lifetime \citep[e.g.,][]{Stelzer2013} and this results in an underestimate of the mass loss.

The relative Ly$\alpha$/Bolometric flux is roughly 1 order of magnitude higher for this M dwarf than it is for the Sun, which has grave implications for photolytic destruction of molecules in planets around M dwarfs of this mass. Even when considering the EUV spectrum of GJ 1132 \citep[calculated with methods described in][]{Youngblood2016} and the EUV flux of the Sun \citep{Zhitnitsky}, we find that GJ 1132 emits 6x as much EUV flux (relative to F$_{\rm{bol}}$) as the Sun.

This work leaves us with several possible pictures of the atmosphere of GJ 1132b:
\begin{itemize}

    \item The real atmospheric loss rates may be comparable to these upper limits, or they may be much less, which leaves us with an open question about the atmosphere and volatile content of GJ 1132b. There could be some loss, but below the detection limit of our instruments.
    \item If there is a neutral hydrogen envelope around GJ 1132b, then this super-Earth is actively losing water driven by photochemical destruction and hydrodynamic escape of H. The remaining atmosphere will then be rich in oxygen species such as O$_2$ and the greenhouse gas CO$_2$.
    \item GJ 1132b could be Mars-like or Venus-like, having lost its H$_2$O long ago, with a thick CO$_2$ and O$_2$ atmosphere remaining, or no atmosphere at all. We posit that this is the most likely scenario, and thermal emission observations with JWST \citep{Morley2017} would give further insight to the atmospheric composition of GJ 1132b.
    \item There might be a giant cloud of neutral hydrogen around GJ1132b based on the \texttt{EVE} simulations, which is undetectable because of ISM absorption. However, if there are other volatiles in the atmosphere we could detect this cloud using other tracers such as carbon or oxygen with HST in the FUV, or helium \citep{2018Natur.557...68S} with ground-based high-resolution infrared spectrographs \citep[see][]{2018Sci...362.1384A, 2018Sci...362.1388N} or with JWST.
\end{itemize}

GJ 1132b presents one of our first opportunities to study terrestrial exoplanet atmospheres and their evolution. While future space observatories will allow us to probe longer wavelength atmospheric signatures, these observations are our current best tool for understanding the hydrogen content and possible volatile content loss of this warm rocky exoplanet. 

\section{acknowledgments}

This work is based on observations with the NASA/ESA Hubble Space Telescope obtained at the Space Telescope Science Institute, which is operated by the Association of Universities for Research in Astronomy, Incorporated, under NASA contract NAS5-26555, and financially supported through proposal HST-GO-14757 through the same contract. This material is also based upon work supported by the National Science Foundation under grant AST-1616624. This publication was made possible through the support of a grant from the John Templeton Foundation. The opinions expressed in this publication are those of the authors and do not necessarily reflect the views of the John Templeton Foundation. ZKBT acknowledges financial support for part of this work through the MIT Torres Fellowship for Exoplanet Research. JAD is thankful for the support of the Heising-Simons Foundation. This project has been carried out in part in the frame of the National Centre for Competence in Research PlanetS supported by the Swiss National Science Foundation (SNSF). VB and DE acknowledge the financial support of the SNSF. ERN acknowledges support from the National Science Foundation Astronomy \& Astrophysics Postdoctoral Fellowship program (Award \#1602597). This project has received funding from the European Research Council (ERC) under the European Union's Horizon 2020 research and innovation programme (project Four Aces; grant agreement No 724427). This work was also supported by the NSF GRFP, DGE 1650115.

\bibliography{gj1132bbiblio.bib}

\end{document}